**Ferroelectric domain nucleation and switching pathways in hafnium oxide.**

Sebastian Calderon V[1], Samantha T. Jaszewski[2], Kyle P. Kelley[3], Jon F. Ihlefeld[2,4], and Elizabeth Dickey[1]

[1] Department of Materials Science and Engineering, Carnegie Mellon University, Pittsburgh, Pennsylvania 15213, USA

[2] Department of Materials Science and Engineering, University of Virginia, Charlottesville, Virginia 22904, USA

[3] Center for Nanophase Materials Sciences, Oak Ridge National Laboratory, Oak Ridge, TN 37831, USA

[4] Charles L. Brown Department of Electrical and Computer Engineering, Charlottesville, Virginia 22904, USA

Nanoscale ferroelectrics that can be integrated into microelectronic fabrication processes are highly desirable for low-power computing and non-volatile memory devices. However, scalable novel ferroelectric materials, such as hafnium oxide ($HfO_2$), remain in a state of development, and a clear understanding of the effects of relevant compositional and processing parameters to control their ferroelectric properties and the actual polarization switching mechanisms are still under investigation. One key fundamental knowledge gap is the polarization switching pathway in ferroelectric hafnia. To further our fundamental understanding of domain nucleation and switching, we have studied polarization switching pathways in $HfO_{2-x}$ thin films in real-time at the atomic scale using transmission electron microscopy. We employed differential phase contrast imaging that allows for the acquisition of both hafnium and oxygen atomic column signals and facilitates the observation of relative movement of atomic columns between both sublattices. Our results demonstrate that the switching pathway involves a transient tetragonal-like local structure, as oxygen ions shift in locations and remain within their parent hafnium polyhedra.

Since the discovery of ferroelectricity in $HfO_2$-based thin films[1], intensive research has been concentrated on understanding the phase stabilizing mechanisms and origin of ferroelectric behavior in these materials. While the stabilizing mechanisms are becoming clearer, the transformations allowing for polarization reversal remain largely experimentally unknown. $HfO_2$ thin films have been shown to co-exist in different crystalline structures, including a thermodynamically stable monoclinic $P2_1/c$ phase, a tetragonal $P4_2/nmc$ phase, and a polar orthorhombic $Pca2_1$ phase. More recently, an antipolar orthorhombic phase has been also demonstrated[2], where the arrangement of oppositely polarized $Pca2_1$ orthorhombic unit-cells is evidenced when observed along the [100] zone axis, which doubles the unit cell and leads to the *Pbca* symmetry. Ferroelectricity in $HfO_2$ compounds is associated with the orthorhombic phases and is caused by the ability of oxygen ions to move in the structure relative to the hafnium sublattice, which occurs either in the polar or antipolar phases. **Figure 1A** shows schematically two polar orthorhombic $Pca2_1$ unit cells viewed along the [100] direction with inverted polarization. The oxygen O1 ions are positioned on a plane that is equidistant from the adjacent hafnium ion planes in the [001] direction. The O2 ions, however, are displaced toward one of the hafnium ion planes, which leads to the non-centrosymmetric structure. The resulting polarization magnitude and direction are assigned by calculating the polarization from a centrosymmetric reference phase, based on the modern theory of polarization[3], as described in the methods section. From this theory, it is important to note that the polarization , both sign and magnitude, can change depending on the switching pathways or the centrosymmetric phase used for the calculation of the polarization, as shown in **Figure S1** (see supplementary information)[4,5].

In traditionally studied perovskite ferroelectrics, such ambiguity does not exist. The switching pathways have been explicitly identified, mainly because the ion responsible for switching the polarization is only minimally displaced within the polyhedra defined by the surrounding cations and anions. However, in novel ferroelectrics such as $HfO_2$, $Hf_{1-x}Zr_xO_2$ (HZO), $Al_{1-x}B_xN$ (AlBN), $Al_{1-x}Sc_xN$ (AlScN), and $Zn_{1-x}Mg_xO$ (ZMO) the switching mechanisms are still under investigation. Recent research from Calderon *et al.*[6] experimentally demonstrated the formation of an antipolar arrangement of wurtzite AlBN during domain nucleation and switching via in-situ scanning transmission electron microscopy (STEM), providing new insight into the polarization reversal mechanisms in wurtzites.

The switching pathways in fluorite structured $HfO_2$, HZO, and $ZrO_2$ have been studied theoretically and have been shown to be complex and depend on the strain state of the films, vacancies, type of dopants, etc. Several pathways for polarization inversion have been proposed by first-principles calculation and nudged elastic band simulations[4,5,7–10]. These pathways are differentiated by the displacements of the oxygen ions relative to the hafnium sublattice and by the particular transient non-polar centrosymmetric state traversed during switching. These transient states are a tetragonal $P4_2/nmc$ phase or an orthorhombic $Pbcm$ phase. Polarization inversion through these two transient states have been reported to have different barrier energies, one being more favorable than the other, depending on the mechanisms of the polarization switching (e.g. uniform polarization or domain-wall motion) and the chemistry of the films (e.g. $HfO_2$, $ZrO_2$, $Hf_{1-x}Zr_xO_2$ or $Hf_{1-x}La_xO_2$, etc).

These two different switching pathways have been predicted to exhibit opposite sign polarization responses, as illustrated in **Figure S1**. Some studies have proposed to use the sign of the piezoelectric coefficient ($d_{33}$) as a key parameter to experimentally determine the actual switching pathway occurrent in the material[7]. However, this coefficient is highly dependent on strain and film texture[11], showing variation in the sign by altering the texture of the film (see **Figure S2 and S3**). Thus, experimental understanding the atomic-scale mechanisms and pathways for domain nucleation and polarization inversion remains elusive and has not yet been experimentally realized, but is important for future material design, and imperative to properly calculate the polarization magnitude and to utilize the material in devices that rely upon the switching and electromechanical responses.

In this report, we induce polarization switching in oxygen-deficient $HfO_{2-x}$ films in which both polar and antipolar arrangements (e.g. collocated $Pca2_1$ and $Pbca$ phases) coexist. The polarization event is triggered by electron irradiation of the sample demonstrating the existence of one of the switching pathways previously calculated in the literature.

**Figure 1B and 1C** show annular dark field (ADF) and differentiated differential phase contrast (dDPC) images obtained simultaneously in one grain. ADF images do not provide information about the oxygen sublattice position and, hence, cannot distinguish polarization direction in a grain. However, using dDPC, it is possible to observe O and Hf atomic columns, allowing for the identification of the spontaneous polarization, when observed along the [100] zone axis of the $Pca2_1$ polar phase. Although along this projection, the O2 atomic columns (see **Figure 1**A) have a projected distance of 0.72 Å, which is very close to the limit of the STEM resolution, the relative position between the O2 and the Hf sublattice allows a simple determination of the polarization along the $c$-axis, $P_z$.

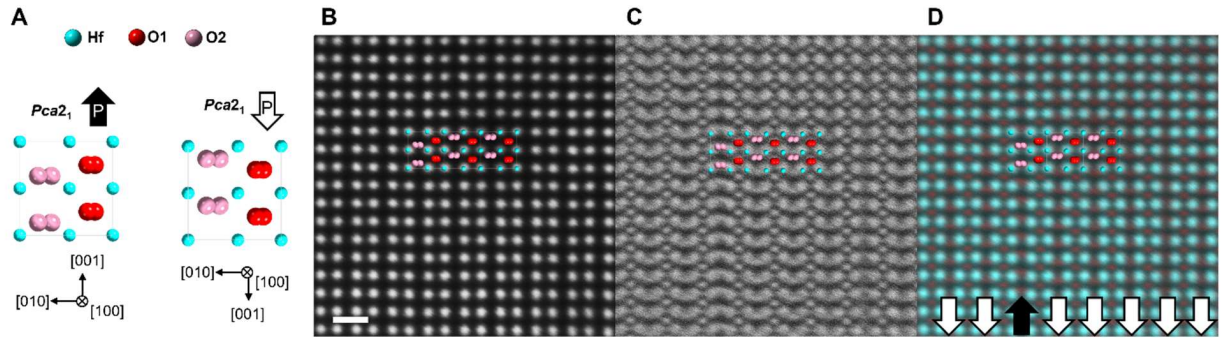

**Figure 1. Local polarization determination via STEM. (A)** Schematic of the orthorhombic $Pca2_1$ structure viewed along the [100] direction for two polarization directions. **(B)** ADF-STEM image of a $HfO_{2-x}$ grain viewed along the [100] for the $Pca2_1$ orthorhombic phase. **(C)** dDPC-STEM image for the same region shown in A. **(D)** overlay of ADF and dDPC- STEM images to highlight the oxygen columns (red) over the Hf atomic columns (light blue). Scale bar is 0.5 nm.

Polarization switching in these materials has been demonstrated by electrical measurements using P-E loops[1], piezoresponse force microscopy[12–15], and some studies using scanning electron microscopy[2,16]. In the latter case, high angle annular dark-field (HAADF), and annular bright-field (ABF) image modalities have been used to image ex-situ switching, showing a reversible transformation between ferroelectric and antiferroelectric orthorhombic phases[2]. In-situ studies, where the electron probe is used to trigger the polarization switching, to our knowledge, have been limited to HAADF, demonstrating ferroelastic behavior in these materials[16]. However, none of these studies offer information about the switching pathways to attain polarization reversal.

Using STEM imaging, we generate a polarization inversion event in undoped hafnium oxide ($HfO_{2-x}$[17,18]) by exposing the materials to a highly energetic electron beam and taking advantage of the high oxygen deficiency, which is expected to facilitate polarization switching in these materials[19,20]. The in-situ switching process is shown in supplementary information (**Movie S1**) and summarized in **Figure 2**, where the primary unit cells, showing domain nucleation associated with polarization inversion, are highlighted in yellow. The figure shows a set of frames during the polarization reversal, in both dDPC (**Figure 2A**) and ADF (**Figure 2B and 2C**) image modalities. dDPC shows the location of both hafnium and oxygen atomic columns, facilitating the observation of the pathway taken by the oxygen atoms to reverse polarization. A qualitative analysis of the images is performed to determine the polarization of each unit cell, by considering the relative position of the Hf vs O atomic columns as shown in **Figure S4**. The results indicate a nucleation event occurring at 99 s in the topmost left unit cells of the frame, where the atomic columns representing the O2 oxygen ions move from the bottom of the hafnium polyhedra towards the top of the frame (along the *c*-axis). Polarization switching propagates down along the same axis to four unit cells more, and then back-switching to the initial configuration is observed after 251 s.

Although ADF does not provide information about the oxygen locations, it offers an accurate determination of Hf-Hf interatomic projected distances to evaluate small changes in the local lattice parameters. ADF images are overlayed with red, white, and blue lines, indicating the projected interatomic distance between Hf-Hf atoms along the *b*-axis (middle panel) and *c*-axis (bottom panel). The images depict an alternating pattern from short (blue) to long (red) atomic spacing along the *b*-axis, characteristic of the interplanar Hf-Hf distances for the *Pca2*$_1$ phase projected along the [100] direction. No significant changes of the local *b*-axis lattice parameters are observed during the switching event. However, slightly larger Hf-Hf *c*-axis interplanar distances are observed at the left-top corner of the image, where the polarization inversion occurs.

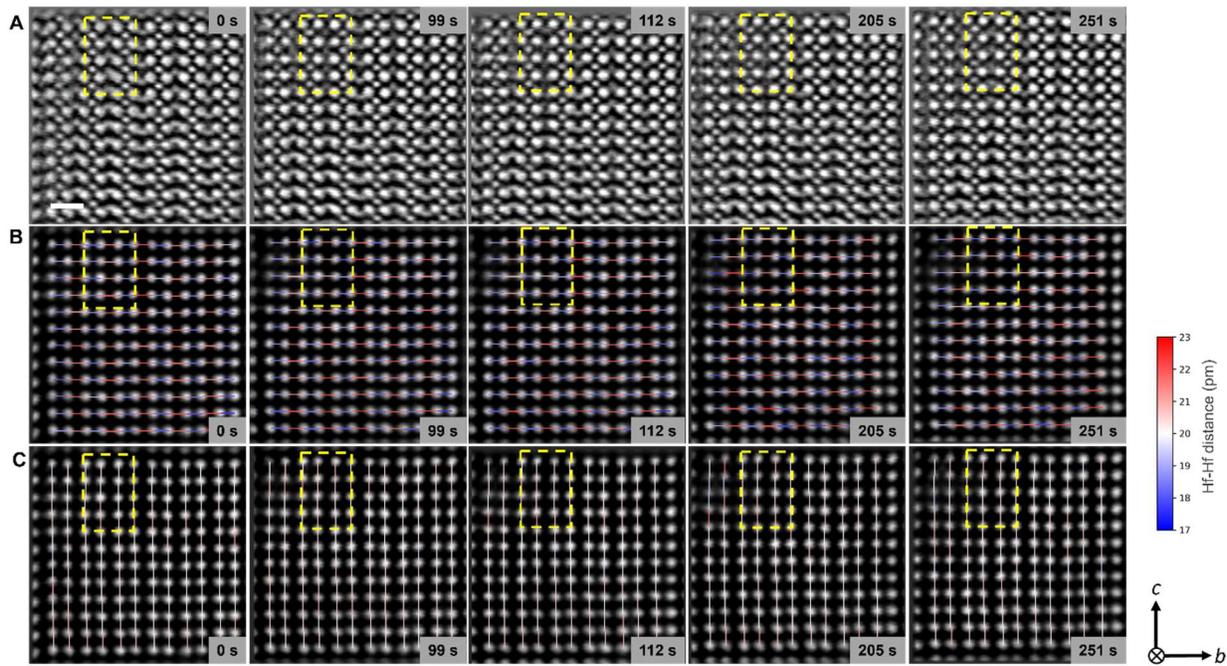

**Figure 2. In-situ polarization switching. (A)** dDPC-STEM image of a HfO$_{2-x}$ grain viewed along the [100] for the Pca2$_1$ orthorhombic at 0, 99, 112, 205 and 251 s. ADF-STEM images for the same region in A, showing Hf-Hf projected interplanar distances **(B)** along the *b*-axis and **(C)** along the *c*-axis. Scale bar is 0.5 nm.

**Figure 3A** shows a set of three isolated dDPC images of two unit cells, where a clear identification of the switching pathway is observed. The initial frame at 0 s shows the oxygen atomic columns (O2) located closer to the bottom hafnium atomic columns in the unit cell, as schematically presented in the atomic model. After 99 s, the oxygen atoms move up in the structure, passing through the center of the Hf polyhedra, indicating a transient distorted tetragonal-like state. Note that the O1 atomic columns maintain their coordination and do not move during the nucleation and switching process. Additionally, as previously highlighted, the *b*-axis Hf-Hf atomic distances during the switching process do not alter significantly, maintaining the long-short pattern in the interatomic distances along the *b*-axis during the entire transition, indicating that the structure is not completely transformed into a tetragonal lattice. As time progresses, oxygen atoms continue the same trajectory, until a complete reverse of the polarization is obtained. A detailed analysis of the Hf-Hf interatomic distances as a function of time is presented in **Figure 3B**. Light blue circles represent the Hf-Hf interplanar distances along the *b*-axis across the full image, while dark-blue circles correspond to the same atomic column distance for the topmost unit cell that switches during the experiment. Similarly, pink circles represent the Hf-Hf interplanar distances along the *c*-axis across the entire field of view, while red circles depict the same distance for the unit cell where polarization inversion occurs. On average, this unit cell has a 1.8% tensile strain along the *c*-axis and 1.5% compressive along the *b*-axis compared to the unit-cells in the entire field of view. However, no significant systematic variation is observed during the switching process, indicating that the Hf sublattice maintains the initial structure before, during, and after the switching event.

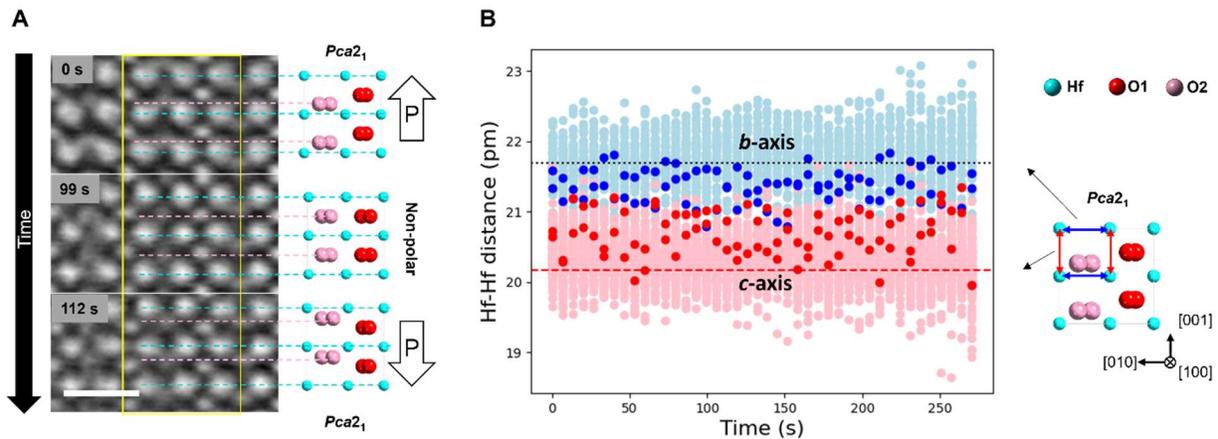

**Figure 3. Transient state structural characterization. (A)** dDPC-STEM image at 0, 99 and 112 s of a $HfO_{2-x}$ grain viewed along the [100] for the $Pca2_1$ orthorhombic phase at 0, 99, 112, 205 and 251 s. **(B)** Hf-Hf projected interplanar

distances along *b*-axis (light/dark blue circles) and c-axis (red/pink circles). The dotted lines in B indicate the Hf-Hf distances for a reference orthorhombic $Pca2_1$ structure. Scale bar is 0.5 nm.

Our work shows experimental evidence for a switching pathway through a transient non-polar structure that resembles a strained tetragonal phase, for highly oxygen-deficient $HfO_{2-x}$ films. It is important to note that despite our results showing the existence of such switching pathway, it does not discard additional pathways, which may become energetically favorable from the existence of strain fields, chemical heterogeneities (e.g., $Hf_{1-x}Zr_xO_2$ or $Hf_{1-x}La_xO_2$, etc.), or other types of defects. The experimental evidence confirms the existence of one of the switching mechanisms predicted by nudged elastic band calculations[4]. Importantly, confirmation of the switching mechanism allows for new insight into domain nucleation, which necessarily requires local polarization switching. This new fundamental knowledge of the switching pathway also may enable development of means to reduce the coercive fields, for example through defect or strain engineering. Such developments would provide larger margins between switching and breakdown voltages and increases in device endurance necessary to realize ferroelectric hafnia's widespread use in low energy computing technologies.

**Materials and Methods**

A 20 nm $HfO_{2-x}$ film was deposited on 85 nm thick TaN bottom electrodes on (001) silicon substrates by reactive High-power impulse magnetron sputtering (HiPIMS) with a Starfire Impulse HiPIMS power module in conjunction with a dc power supply to deliver square voltage pulses with a duration of 110 μs, a frequency of 200 Hz, and a magnitude of -700 V. These parameters represent a duty cycle of 2.0 % and per-pulse peak plasma power density of 600 W/cm². Following each negative high-power pulse, a +100 V positive pulse of length 200 μs was applied to the target. Argon was used as the sputter gas, and $O_2$ gas was used as the reactive gas. A gas flow of 15.00 sccm of argon with 1.23 sccm of $O_2$ was controlled by electronic mass flow controllers, and the total process pressure was controlled by a conductance flow valve located in front of the turbomolecular pump, reaching oxygen concentrations in the background gas of 7.6%. The substrate was grounded, continuously rotated, and not intentionally heated during deposition. Following deposition, the film was annealed in a pure argon atmosphere (99.999% purity) in a rapid thermal processor, Allwin 21, Heatwave 610, at 750 °C for 30 s. Additional material properties are provided in Jaszewski, *et al.*[17]

The area detector X-ray diffraction measurement was completed using a Bruker APEXII Duo Single Crystal X-Ray diffractometer in a reflection geometry with a collimated Cu Kα radiation source, a fixed omega angle of 15°, and an APEXII CCD area detector. Magnesium oxide powder was adhered to the film surface as a height reference standard.

TEM samples were prepared by a cross-sectional wedge mechanical polishing method. To decrease the damage from the mechanical polish, final thinning was achieved by argon ion polishing using a Gatan PIPS II instrument with sequential accelerating voltages of 2 kV, 1 kV, and 500 V. Simultaneous annular dark field (ADF) and differential phase contrast (DPC) images were acquired using a ThermoFisher Themis microscope operated at 200 kV, with a probe convergence angle of 17.9 mrads and acceptance angles between and 46 – 200 mrads (ADF) and 11- 43 mrads (DPC). A set of fast-dwell-time (200 ns) frames were collected, and drift corrected using a cross-correlation method to avoid distortions in the images. In-situ images were acquired using a continuous acquisition over approximately 5 min at a 5 μs dwell-time for 256 x 256 frames. The images reported were averaged every 20 frames to improve the signal-to-noise ratio. Differentiated differential phase contrast (dDPC) images were obtained as the gradient of the DPC images[21].

The projected polarization was calculated by assuming two different centrosymmetric phases, namely tetragonal *P4₂/nmc* and orthorhombic *Pbcm*. The displacement of oxygen anions and hafnium cations with respect to the tetragonal or orthorhombic *Pbcm* reference unit cells was input into eq. 1, which is based on the modern theory of polarization[3].

$$\Delta P = \frac{e}{\Omega} \sum_i Z_i^* \vec{r_i} \quad (1)$$

$\Delta P$ is the change in polarization, $e$ is the positive electronic charge, $\Omega$ is the unit cell volume, $Z^*$ is the Born effective charge[5], $\vec{r}$ is the displacement of the ions from one state to another, and the subindex $i$ iterates along all the atoms in the structure. Note that the displacement of the Hf cations and the O1 anions can be assigned to $\vec{r} = 0$, and only the O2 ions are displaced ($\vec{r}_{O2}$), as shown in **Figure S1**. In addition, since the change in polarization is calculated with respect to a centrosymmetric reference structure, the resulting $\Delta P$ can be interpreted as the spontaneous polarization of the polar phase considering the same displacement for each O2 ion. Eq. 1 can be simplified as $P_s = \frac{4eZ_{O2}^* \vec{r}_{O2}}{\Omega}$, where $e$ is the positive electronic charge, $\Omega$ is the unit cell volume, $Z_{O2}^*$ is the Born effective charge for O anions and $\vec{r}_{O2}$ is the O2 ion displacement from the reference structure.

All piezoresponse force microscopy (PFM) measurements (see **Figure S3**) were conducted with the same Budget Sensor Multi75E-G AFM probe (Cr/Pt coated, ~3 N/m) using an Oxford Instruments Cypher atomic force microscope. Band excitation PFM was achieved via connection to an arbitrary wave generator and data acquisition electronics powered by a National Instruments fast DAQ card. A custom software program was employed to generate the probing signal and record local piezoresponse and hysteresis loops. Band excitation PFM measurements were performed at a frequency range of 300-400 kHz and amplitude of 1 V. Simple harmonic oscillator fits were subsequently applied to the collected spectra to extract amplitude and phase information at the resonance frequency.

**Acknowledgments**

This material is based upon work supported by the Center for 3D Ferroelectric Microelectronics (3DFeM), an Energy Frontier Research Center funded by the U.S. Department of Energy, Office of Science, Office of Basic Energy Sciences Energy Frontier Research Centers program under Award Number DE-SC0021118. S.T.J. acknowledges support from the U.S. National Science Foundation's Graduate Research Fellowship Program under grant DGE-1842490. The authors acknowledge the use of the Materials Characterization Facility at Carnegie Mellon University supported by grant MCF-677785. This work utilized a Bruker D8 Venture instrument, which was acquired under Award CHE-2018870 from the U.S. National Science Foundation's Major Research Instrumentation program.

# Supplementary information

**Atomic scale observation of ferroelectric switching pathways in ferroelectric hafnium oxide.**

Sebastian Calderon V[1], Samantha T. Jaszewski[2], Kyle P. Kelley[3], Jon F. Ihlefeld[2,4], and Elizabeth Dickey[1]

[1] Department of Materials Science and Engineering, Carnegie Mellon University, Pittsburgh, Pennsylvania 15213, USA

[2] Department of Materials Science and Engineering, University of Virginia, Charlottesville, Virginia 22904, USA

[3] Center for Nanophase Materials Sciences, Oak Ridge National Laboratory, Oak Ridge, TN 37831, USA

4 Charles L. Brown Department of Electrical and Computer Engineering, Charlottesville, Virginia 22904, USA

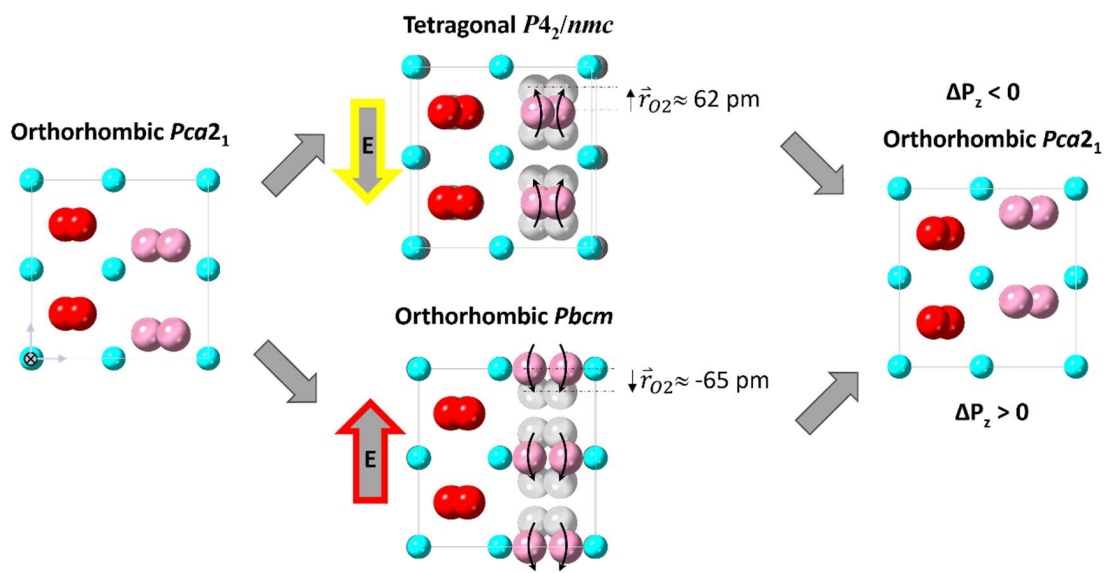

**Figure S1. Alternative switching pathways.** Two centrosymmetric phases have been identified to allow polarization switching in $HfO_2$ by using DFT calculation[4]. A tetragonal $P4_2/nmc$ and an orthorhombic *Pbcm*. Both paths require opposite oxygen pathways to reverse the polarization, which results in opposite sign of the polarization.

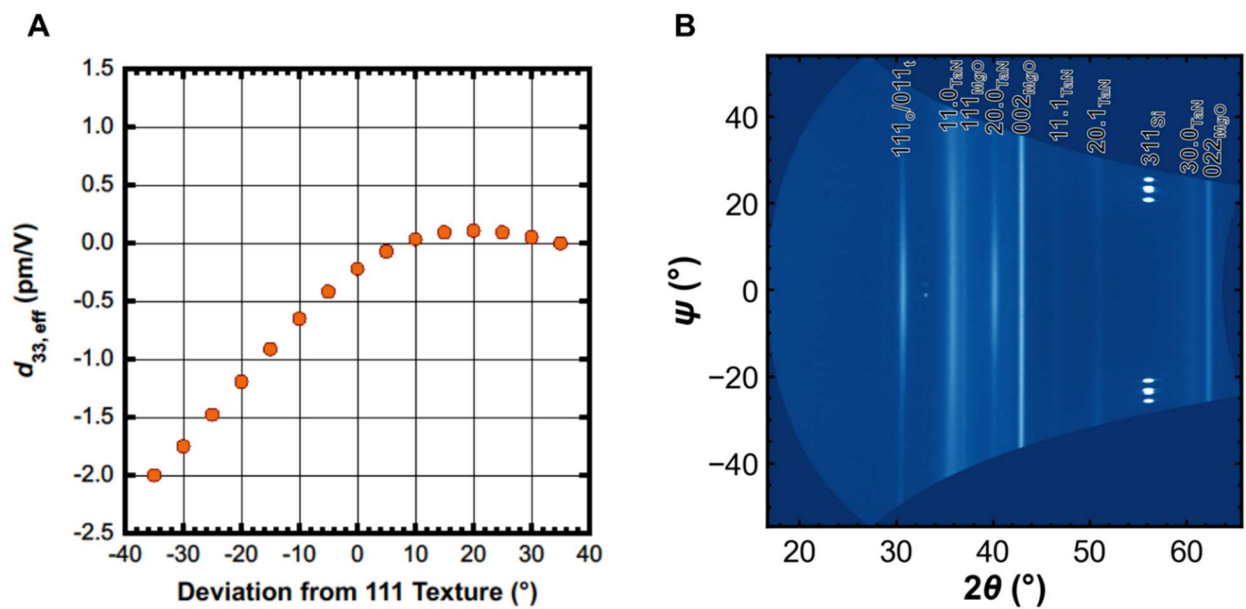

**Figure S2. Piezoelectric effective coefficient $d_{33, eff}$ as a function of the texture of the films.** Using the equation reported in [11], it is shown that the sign of $d_{33, eff}$ can vary depending on the texture of the films. (**A**) show the calculated value for $d_{33, eff}$ changing the texture of the films from a pure 111 texture to ± 40 degrees. The results show negative to positive value depending on the degree of texture. (**B**) 2D X-ray diffraction scan for the samples presented in this manuscript, showing a 111 preferential orientation.

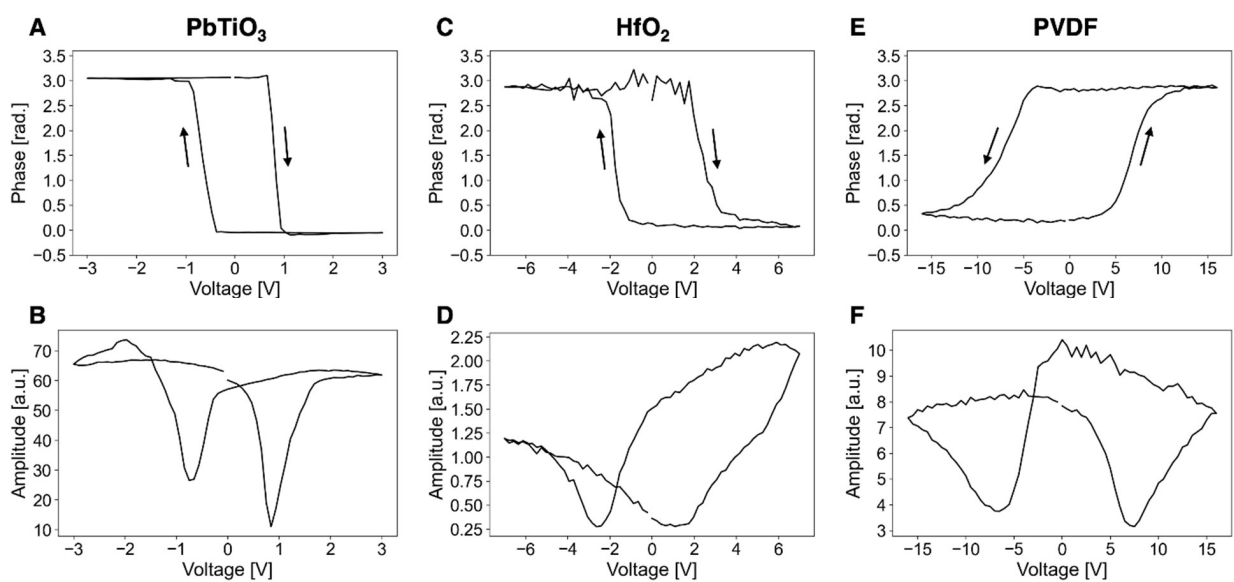

**Figure S3. Band excitation piezoresponse force microscopy.** Average band excitation piezoresponse force microscopy off-field phase (top panels) and amplitude (bottom panels) hysteresis loops acquired on (A,B) PbTiO3, (C,D) HfO2, and (E,F) PVDF. Note, the sign of the piezoelectric coefficient determines PFM phase switching handedness, i.e. clockwise and counter clockwise phase rotation for positive and negative d33, respectively. PbTiO3 has a positive d33, while PVDF has negative d33.

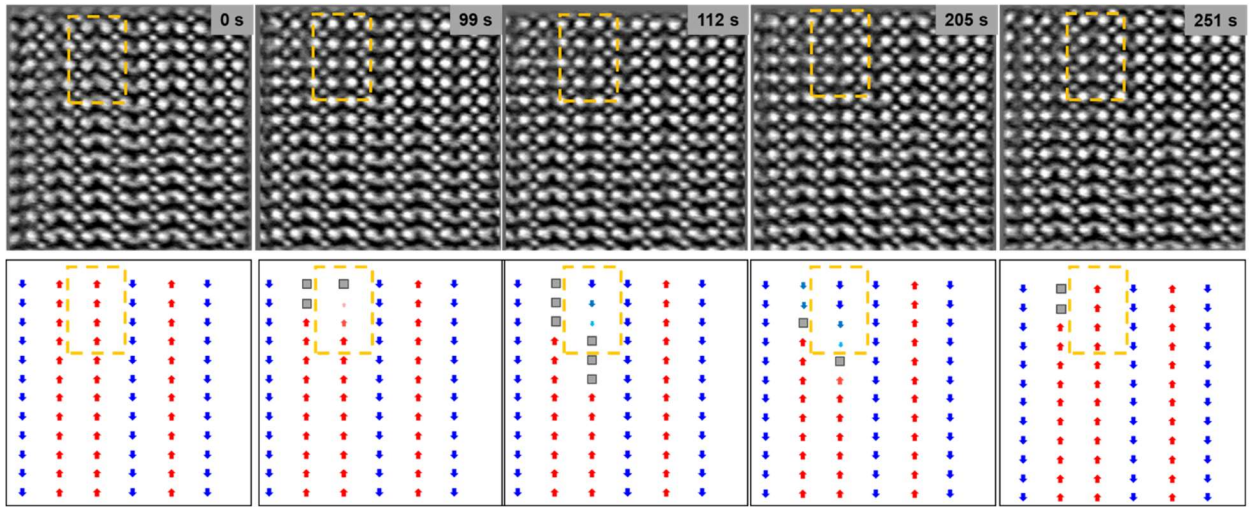

**Figure S4. Experimental nucleation and switching pathways.**